\documentclass[a4paper,11pt]{article}
\usepackage{jinstpub} 

\usepackage[T1]{fontenc} 

\usepackage[dvipsnames,table,xcdraw]{xcolor}

\title{\boldmath Data-driven detector signal characterization with constrained bottleneck autoencoders}

\author[a,1]{C.Jes\'us-Valls,\note{Corresponding author.}}
\author[a]{T.Lux,}
\author[b]{F.S\'anchez}

\affiliation[a]{Institut de F\'isica d’Altes Energies (IFAE) - The Barcelona Institute of Science and Technology (BIST), Campus UAB, 08193 Bellaterra (Barcelona), Spain}
\affiliation[b]{ University of Geneva, Section de Physique, DPNC, 1205 Genève, Switzerland}

\emailAdd{cesar.jesus@cern.ch}

\pdfoutput=1 
\usepackage[dvipsnames,table,xcdraw]{xcolor}
\usepackage{graphicx}

\graphicspath{{Figures/}}

\begin{document}

\abstract{A common technique in high energy physics is to characterize the response of a detector by means of models tuned to data which build parametric maps from the physical parameters of the system to the expected signal of the detector. When the underlying model is unknown it is difficult to apply this method, and often, simplifying assumptions are made introducing modeling errors. In this article, using a waveform toy model we present how deep learning in the form of constrained bottleneck autoencoders can be used to learn the underlying unknown detector response model directly from data. The results show that excellent performance results can be achieved even when the signals are significantly affected by random noise. The trained algorithm can be used simultaneously to perform estimations on the physical parameters of the model, simulate the detector response with high fidelity and to denoise detector signals.}

\maketitle

\section{Introdution}
In high energy physics (HEP), an usual analysis strategy is that of comparing experimental measurements to synthetic datasets generated by means of a simulator. Given a simulation model, the synthetic data can be tuned by changing the underlying model parameters allowing to compute best fit points and confidence intervals. However, to accurately model a system a precise quantitative knowledge of its behavior is necessary. Sometimes, such a knowledge is only partial, and the discrepancies between the simulated and the measured datasets lead to systematic errors, decreasing the experimental sensitivity.\\
In general, HEP simulations consist of two parts: The propagation of particles in the detector and the detector response. Whereas the same modeling of physics can be used in a large variety of experiments, the detector response is often quite specific. Hence, dedicated characterization studies of a detector are often necessary typically consisting of a multitude of smaller characterizations tasks each describing the behavior of either a sub-detector module or an individual channel. Modeling the detector signals is often non-trivial due to the many effects that can arise in the detector response, either in the signal production (e.g. quenching, de-excitations, etc), in the signal transport (e.g. field non-uniformities, attenuation, diffusion, reflections, spreading, etc) or in the sensor response (e.g. saturation, after-pulsing, cross-talk, etc). Therefore, mismodeling often arises in the detector simulation leading to systematic errors. To improve this, data-driven solutions that characterize the signal with high fidelity are of great interest as they can be used to replace, or improve, the inaccurate detector response models. Due to this, a usual practice is that of adjusting laboratory measurements with functions and use them to characterize the detector signal. Although this is useful for a number of problems, this method is not quite general, as it works only for scalar detector signals. If the signal is instead an array, such as a waveform, the problem grows in complexity. Due to this, simplifying assumptions must often be made which introduce modeling errors. In this article, we present and discuss how constrained autoencoders can be used to learn the underlying unknown detector response models directly from data examples. We illustrate the potential of this method through the study of a toy model producing signal waveforms and we discuss how the trained algorithm can be used to perform parameter estimations, simulate the detector response with high fidelity and to effectively denoise detector signals.

\subsection{Problem definition}
\label{sec:prob_def}
Consider the deterministic response of a detector as a transformation $f$ that maps a set of physics parameters, $\vec{p}$, (e.g. track position, angle, energy deposit, etc) into a set of detector signals $\vec{s}$, namely, 
\begin{equation}
\label{eq:def}
f(\,\,\vec{p}\,\,) \mapsto  \vec{s};\qquad f^{-1}(\,\,\vec{s}\,\,) \mapsto  \vec{p} \,.
\end{equation}
Then, our aim is to learn transformations $\hat{f}$ (and $\hat{f}^{-1}$) as similar as possible to $f$ (and $f^{-1}$) directly from data with the only requirement of knowing the set of $\vec{p}$ playing a role in the system and being able to collect data samples of $\vec{s}$ for known $\vec{p}$ configurations. Noticeably, this single condition can often be satisfied, as output signals $\vec{s}$ can be recorded in the laboratory while determining $\vec{p}$ with the aid of additional instruments. An exemplifying toy model is presented in the following subsection.

\subsection{Toy problem}
We will use a toy transformation $f$ given by:
\begin{align}
\label{eq:f}
\vec{s}(x) &= f(\vec{p}=\{\theta,\eta\})(x)\nonumber\\
&=(1+0.5\theta)\mathcal{M}(-7|1+0.3\theta)+\mathcal{M}(7(\eta-1)|1.2+\theta)
\end{align}
Where the shorthand notation $\mathcal{M}(\mu|c)$ has been used to define a Moyal distribution~\cite{moyal} with location parameter $\mu$ and scale $c$, which describes the energy loss of a charged relativistic particle due to ionization of the medium and approximates the Landau distribution with a systematically lower tail. This model takes two generic input physics parameters $\vec{p}=\{\theta, \eta\}$. To give some physical intuition on the model it has been inspired to resemble the response of a scintillating bar, similar to the signals reported in Ref.~\cite{Korzenev:2021mny}. Under this picture, the first (second) Moyal distribution can be interpreted as accounting for the direct (reflected) scintillating light reaching a photosensor in one end of the bar. Then, one can think of the parameter $\theta$ as accounting for the track angle, related to the light yield, and of $\eta$ as the distance at which the track has crossed the bar with respect to the photosensor, introducing a time delay in the measured signal. If $x$ is interpreted as time, a signal $s(x)$ is measured in the detector at discrete consecutive instances of $x$. In this way, the array of signals $\vec{s}$ constitutes a 1D waveform.\\
Now consider the following situation. The scintillating bar of the previous example can be measured in the laboratory such that a collection of $\vec{s}$ can be recorded. For each of this signals, $\vec{p}$ is also determined with the aid of additional instruments, e.g. additional position sensitive detectors. However, the experimenter is unaware of the true functional form of $f$, which can be in practice arbitrarily complex, but desires to learn $f$, its inverse transformation, or both, out of the collected sets of $\vec{s}$ and $\vec{p}$.\\
In this article, we will use the toy model above to generate data instances of $\vec{s}$ and $\vec{p}$ artificially, reproducing what could be collected in the laboratory. Later, we will pretend that $f$ is unknown, and we will face the problem of learning $\hat{f}$ and $\hat{f}^{-1}$ as close as possible to $f$ and $f^{-1}$ directly from the synthetic data examples.

\section{Autoencoders}
Over the last years deep learning methods have become crucial for modern data analysis and have started to play a important role in HEP. So far, most solutions have been reached through supervised algorithms, particularly applied to classification problems, with deep neural networks at the forefront. Recently, deep generative models have aroused as a novel unsupervised alternative, able to deal with both labeled and unlabeled datasets and to learn sophisticated transformation between spaces of very different dimensionality~\cite{ruthotto2021introduction}. Autoencoders are a particular type of this novel algorithms. In autoencoders (AEs) a real input of arbitrary dimensionality $\mathbb{R}^{A}$ is mapped into a real space output of the same dimensionality by following a set of transformations characterized by a bottleneck architecture. The bottleneck, often referred to as the latent space, has a reduced dimensionality $\mathbb{R}^{B}$, where typically $A\gg B$. As the algorithm is trained to minimize the difference between the input and the output distributions the autoencoder learns a compact representation of the data which is encoded into the latent space. Due to this, the transformation which maps the input into the latent space is named the encoder whereas the transformation which maps the latent space into the output is named the decoder.\\
Despite their recent appearance in the physics literature, AEs have already been studied for several interesting HEP applications. On one hand, AEs can be used to compress data by storing the much lighter latent space representation~\cite{DiGuglielmo:2021ide,Huang:2021ymz,Yoon:2021mim}. On the other hand, AEs average out the noise from different data examples during training and, therefore, they have great potential to denoise signals~\cite{Ai:2019yih,Shen:2019ohi,Erdmann:2019nie}. AE transformations are greatly interesting by themselves. If an AE is trained using non-anomalous signals, for instance produced with a well understood simulation, the AE reconstruction error can later be used to identify outliers and tag them as anomaly candidates. Thus, so far, anomaly detection has been among the most widely explored applications involving AEs in the HEP literature~\cite{Farina:2018fyg,Hajer:2018kqm,Blance:2019ibf,CrispimRomao:2020ucc,Cheng:2020dal}. AEs can also be used to learn signal or background distributions directly from data. In this applications, AEs enable the generation of synthetic data by sampling in the latent space, an option explored in several studies~\cite{Holl:2019xtt,Liao:2021vec,Monk:2018zsb}. However, if the latent space is unconstrained, no physical interpretation can be made of the latent space. Due to this, architectures in where the bottleneck is constrained have been studied in computer science, leading to the proposal of modified AEs, such as the Bounded-Information-Bottleneck AE (BIB-AE)~\cite{voloshynovskiy2019information}. The main difference with a standard AE is that in these modified algorithms additional information in the form of labels is used in the training phase in order to make the latent space interpretable. In HEP, this approach has been first explored recently via an algorithm dubbed the end-to-end Sinkhorn AE. Its use has been explored in the context of producing complex high fidelity simulation outputs for calorimeters by replacing simulators with AEs trained in simulated data~\cite{Buhmann:2020pmy, Buhmann:2021vlp}. This has the main advantage of reducing the time and computational cost of generating simulated events. In this article, we further discuss the potential of latent space constrained autoencoders in HEP. In particular, we focus on a powerful and unexplored application: its use for data-driven detector signal characterization and, even more, how the same trained algorithm can be used not only to generate signals but also to do physical parameter estimations and signal denoising.

\subsection{The constrained bottleneck autoencoder}
\begin{figure}[htb]
\centering 
\includegraphics[width=.9\linewidth]{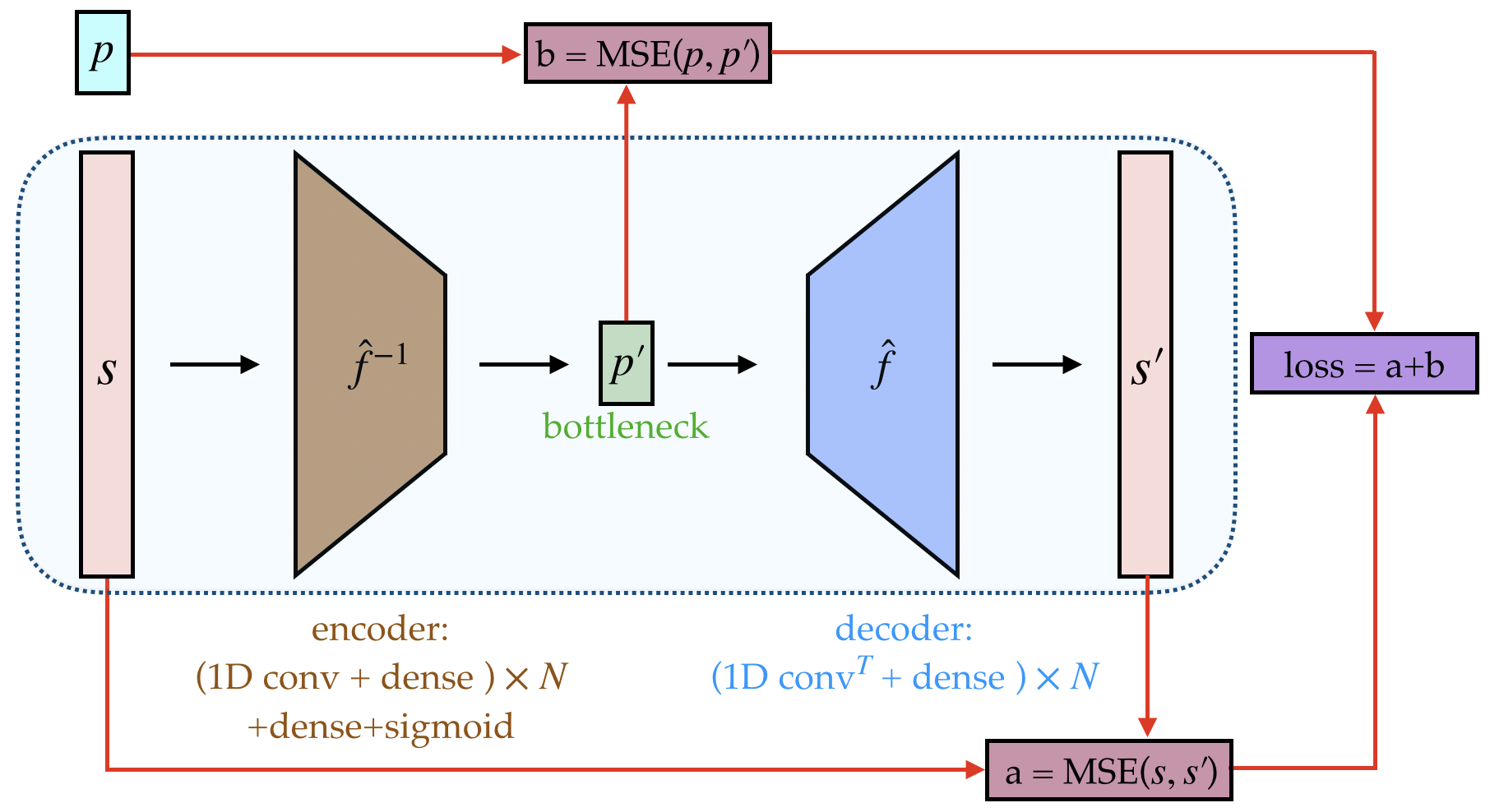}
\caption{\label{fig:cb_ae} Sketch of the constrained bottleneck autoencoder under discussion.}
\end{figure}

We will use a modified AE, similar to the BIB-AE, with simple features which, for generality, will be referred to as a constrained bottleneck autoencoder. A sketch of the algorithm is presented in Figure~\ref{fig:cb_ae}. Its main features are as follows: For training, events consisting of signals $\vec{s}$ associated to physical configuration parameters $\vec{p}$ are given as the input to the algorithm. The algorithm then performs a series of sequential transformations which generate an output signal $\vec{s'}$ with equal dimensionality to $\vec{s}$. The bottleneck $\vec{p'}$ corresponds to the centermost transformation layer and has the same dimensionality as $\vec{p}$. With the former settings, the network weights specifying the encoder and decoder transformations $\hat{f}^{-1}$ and $\hat{f}$, corresponding to our target interest (see Sec~\ref{sec:prob_def}), can be optimized by minimizing the loss over the set of input examples. Such loss consists of the sum of two terms. On one hand, the mean squared error (MSE) between $\vec{s}$ and $\vec{s'}$ which accounts for the similarity of the input and output signals. On the other hand, the MSE between $\vec{p}$ and $\vec{p'}$ which accounts for the similarity between the latent space and the physical configuration parameters.
\subsection{Architecture and implementation}
The former architecture design has been driven by simple principles. In the first place, using convolutional transformation allows to identify patterns arising in the correlations among neighbor signal values. The number of iterations, filters, kernel size and stride were chosen after a set of dedicated architectural tests on the overall performance of the network. In the second place, the use of fully connected layers allows to combine the information from the different convolution filters and provides to the algorithm  enough flexibility in the encoder and decoder transformations. Such flexibility is crucial, as our AE has a constrained bottleneck, and therefore, learning $\hat{f}$ and $\hat{f}^{-1}$ is significantly more challenging than in the unconstrained case. In this sense, stacking N-times the former layers allows to organically increase the complexity of the network, either by increasing the number of filters, the number of repetitions, or both.

\section{Methodology}

\subsection{Dataset generation}
\begin{figure}[ht!]
\centering 
\includegraphics[width=.99\linewidth]{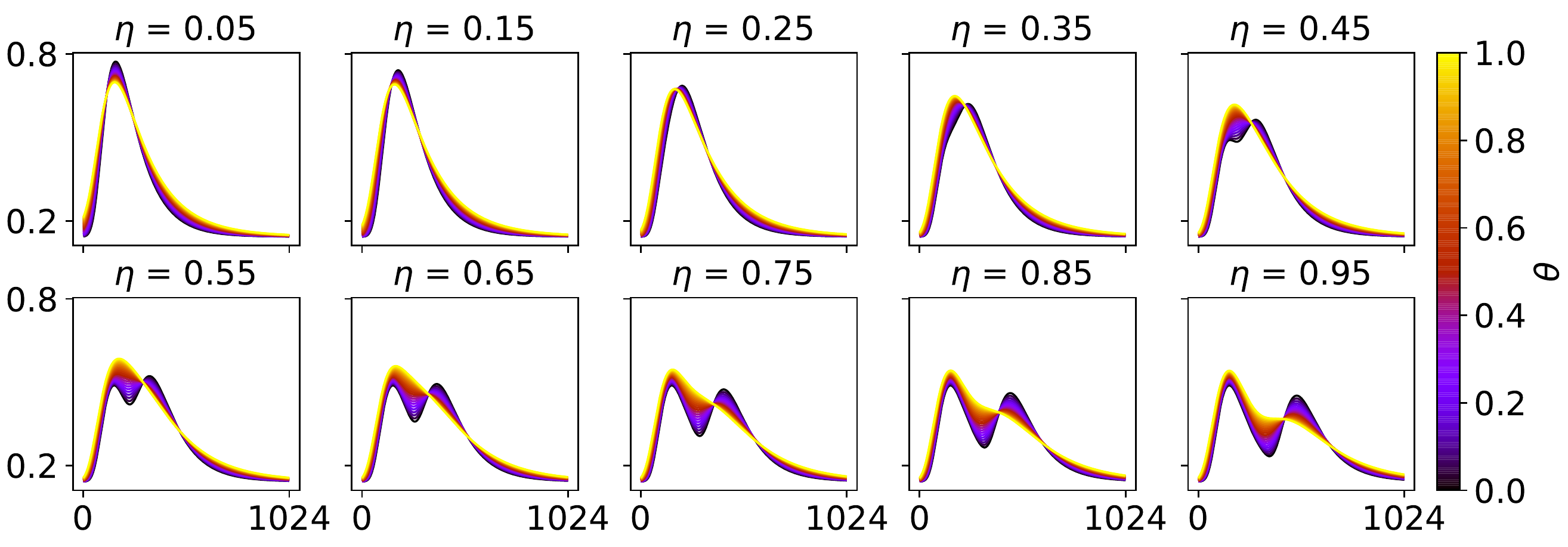}
\caption{\label{fig:signal_shape} Functional normalized shape of $f(\vec{p})$ as described in Eq~\ref{eq:f}. A colormap of $\theta$ for a series of slices of $\eta$ is used to illustrate the dependency on both parameters.}
\end{figure}

\begin{figure}[ht!]
\centering 
\includegraphics[width=.99\linewidth]{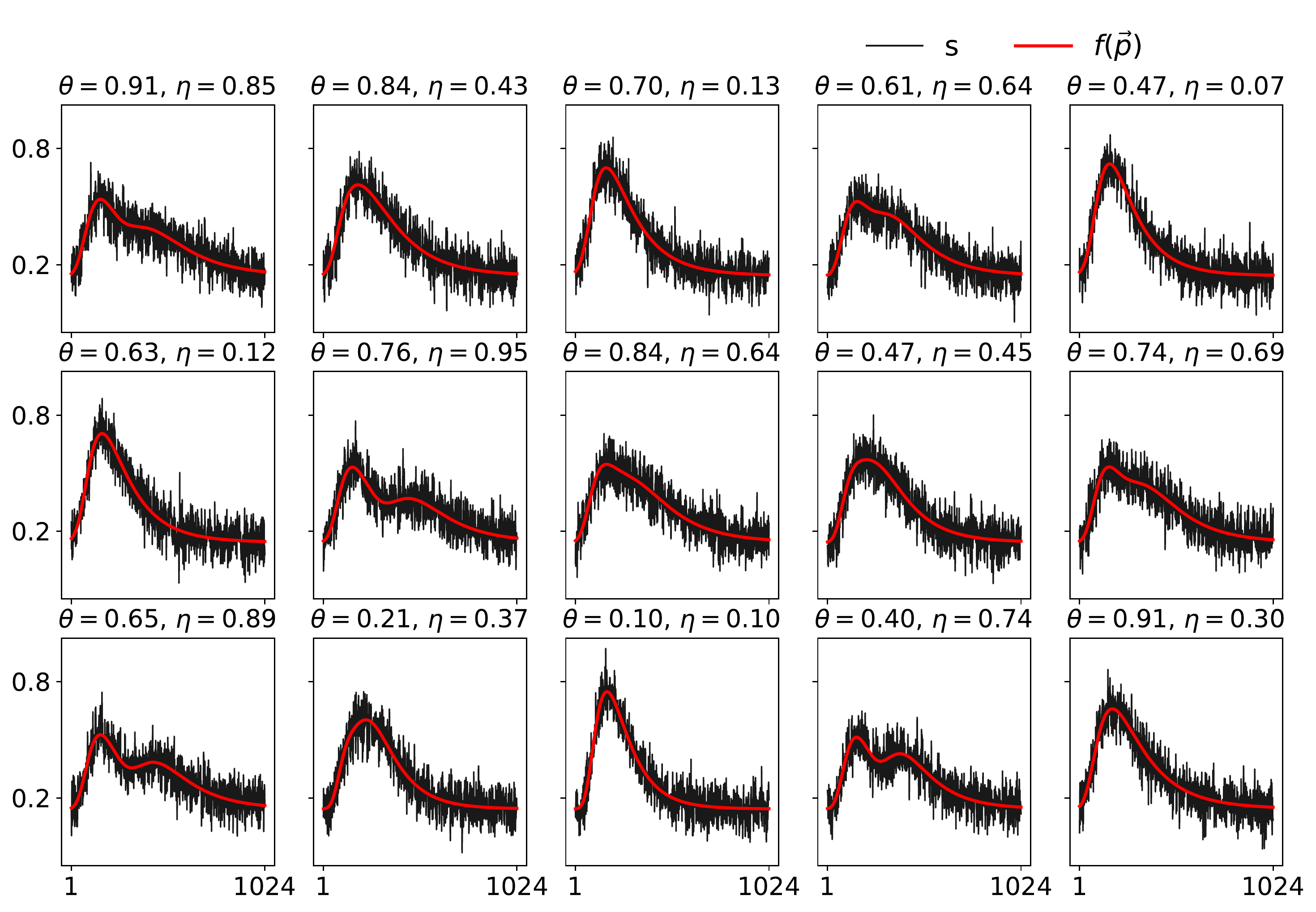}
\caption{\label{fig:noise_examples} Examples of 15 random noisy waveforms in the test dataset compared to its original functional shape, specified by Eq.~\ref{eq:f}.}
\end{figure}

A dataset has been generated by producing a set of 50k statistically independent $\vec{s}$. Each instance of $\vec{s}$ has been generated using uncorrelated flat random values of $\theta\in[0,1]$ and $\eta\in[0,1]$ and by evaluating $f$, as defined in Eq.~\ref{eq:f}, a total of $2^{10}=1024$ equidistant times along an interval of $x\in[-10,10]$. In this way, the signals $\vec{s}$ are waveforms of 1024 values. For training purposes the waveforms have been normalized. This can be achieved in general by applying the change of variable: $\vec{s}=(\vec{s}+\textup{offset})/\textup{scale}$. 
For convenience, we have used offset=0.1 and scale=0.8 such that after adding noise to the waveforms (explained below), the signals are made of values largely contained in the interval [0,1]. An illustration of the normalized waveforms provided by the toy model in Eq.~\ref{eq:f} is presented in Figure~\ref{fig:signal_shape}.
\\In a realistic case, the waveforms would be affected by random fluctuations. To show the robustness of the method to random noise and to illustrate its denoising potential prior to the signal normalization each value of {$\vec{s}$\,} has been varied independently, adding to it the result of random sampling from a Gaussian with the mean at the original signal value and a sigma of 0.05. At the normalization scale=0.8, this level of noise corresponds to $6.25\%$ of the maximum possible signal amplitude.
Example results of a set of 15 randomly chosen instances of $\vec{s}$ compared to their associated $f(\vec{p})$ are presented in Figure~\ref{fig:noise_examples}.

\subsection{Training}
For training and testing the system the dataset of 50k examples has been split into two separate datasets: A training dataset with 40k signals and a test dataset with the remaining 10k events. The autoencoder has been trained for 10 epochs (complete cycles to all the data), using the Adam optimizer~\cite{kingma2014adam} with an initial learning rate of $5\cdot10^{-4}$ and a mini-batch size of 32 (which determines the amount of examples that are processed before updating the network parameters). During training the network weights are optimized by minimizing the loss function, which as earlier anticipated in Figure~\ref{fig:cb_ae}, consists on the addition of the minimum squared error of the signal reconstruction accuracy and the latent space matching accuracy. The training for which the results are reported took about 50~min in a standard computer\footnote{2.3~GHz dual-core Intel Core i5.}.

\section{Results and discussion}
The performance of the trained algorithm can be evaluated for different purposes. In the first hand, the quality of the learned $\hat{f}^{-1}(\vec{s}) \mapsto \vec{p'}$ transformation is relevant to do estimates of the physical parameters $\vec{p}$ using signals as input. On the second hand the quality of the learned $\hat{f}(\vec{p}) \mapsto \vec{s'}$ is relevant to model the detector response and to produce realistic signals $\vec{s}$, e.g. if the algorithm is embedded in a Monte Carlo simulation. Notably, since the algorithm learns from multiple examples the random noise is averaged out, such that $\hat{f}(\vec{p})$ generates noiseless signals $\vec{s'}$, even if it learns from noisy examples $\vec{s}$. Due to this, the transformation $\hat{f}^{-1}(\hat{f}(\vec{s}))\mapsto \vec{s'}$ can be used to denoise signals. We study each of these tasks separately and discuss them below. 

\subsection{Reconstruction of the physical parameters}

\begin{figure}[ht!]
\centering
\begin{minipage}{0.53\linewidth}
\centering 
\includegraphics[width=.99\linewidth]{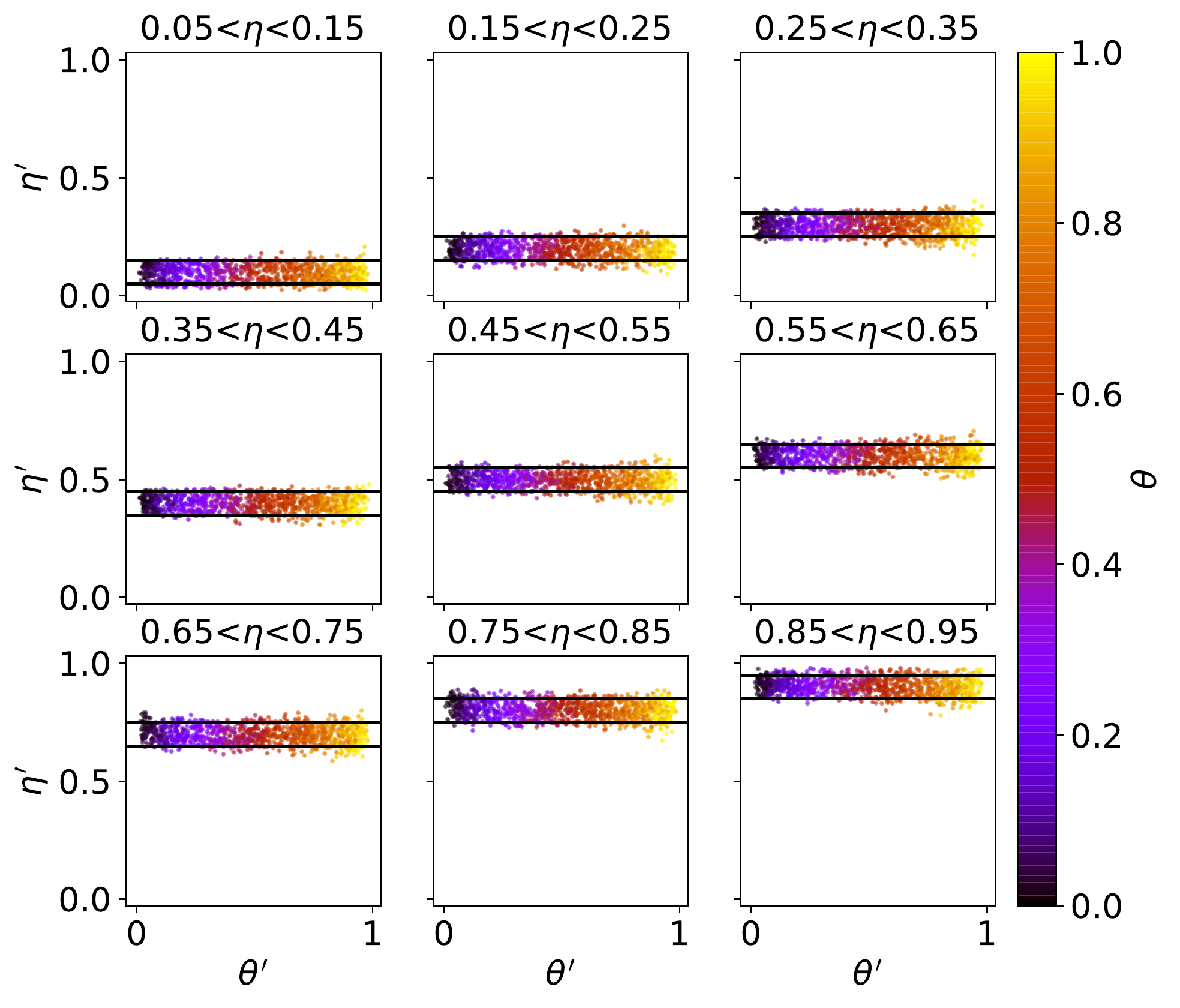}
\caption{\label{fig:latent_space} Distribution of the two reconstructed physical parameters ($\theta'$ and $\eta'$) of the toy model for the whole test dataset. The reconstructed values are shown in slices of true $\eta$ and colored with the true value of $\theta$, allowing to visualize the latent space. The horizontal black lines show the constrains on the true values of $\eta$.}
\end{minipage}\hfill
\begin{minipage}{0.45\linewidth}
\centering 
\vspace{0.3cm}
\includegraphics[width=.99\linewidth]{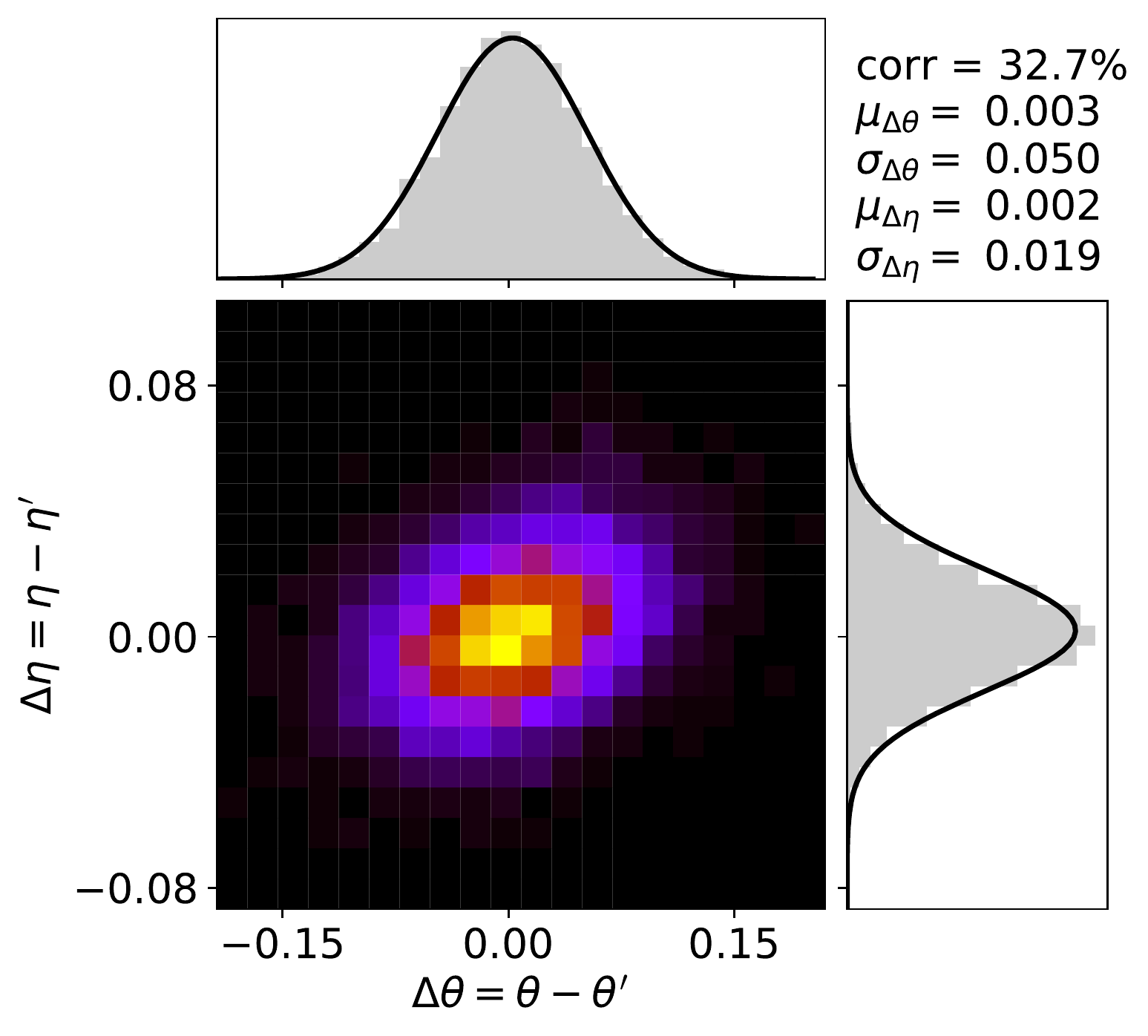}
\vspace{0.cm}
\caption{\label{fig:latent_rec_err} 2D distribution of the reconstruction error for $\theta$ and $\eta$ and its 1D marginal error distributions including Gaussian fits. The correlation of the two errors is also presented.}
\vspace{1cm}
\end{minipage}
\end{figure}

\noindent The distribution of the reconstructed physical parameters, $\vec{p}=\{\theta,\eta\}$, associated to the signals for the whole test dataset of 10k events is Figure~\ref{fig:latent_space}. The results are presented in slices of true $\eta$ and in a colormap of $\theta$ allowing to qualitatively understand the 2D latent space distributions in the whole phase-space. As expected, the encoder learns a transformation that maps signals $\vec{s}$ associated to transformations determined by values of $\vec{p}=\{\theta,\eta\}$ to a latent space $\vec{p'}=\{\theta',\eta'\}$ where the typical distance between $\vec{p}$ and $\vec{p}'$ is small. This is explicitly shown in Figure~\ref{fig:latent_rec_err} where the errors in each of the reconstructed physical parameters is presented as a 2D plot showing pairs of reconstructed $\{\theta',\eta'\}$ and their 1D marginal distributions. Notably, the errors are Gaussian distributed, and show negligible bias and a very narrow standard deviation. The width of the distribution is different for the two parameters. This is expected, as the two parameters affect differently the waveform shape. In particular, looking into Figure~\ref{fig:signal_shape} it is clear that a wide range of $\theta$ can render very similar waveforms, as for instance for small $\eta$ values. From these results on the reconstructed latent space two major conclusions can be extracted. In the first place the algorithm is able to effectively learn the transformation $\hat{f}^{-1}(\vec{s}) \mapsto \vec{p'}$, allowing to do parameter estimations directly from detector signals. In the second place, no outliers are observed and the error distribution is Guassian allowing to easily and reliable associate errors to the parameters estimated by means of this method.

\subsection{Signal generation}
\begin{figure}[ht!]
\centering
\includegraphics[width=.99\linewidth]{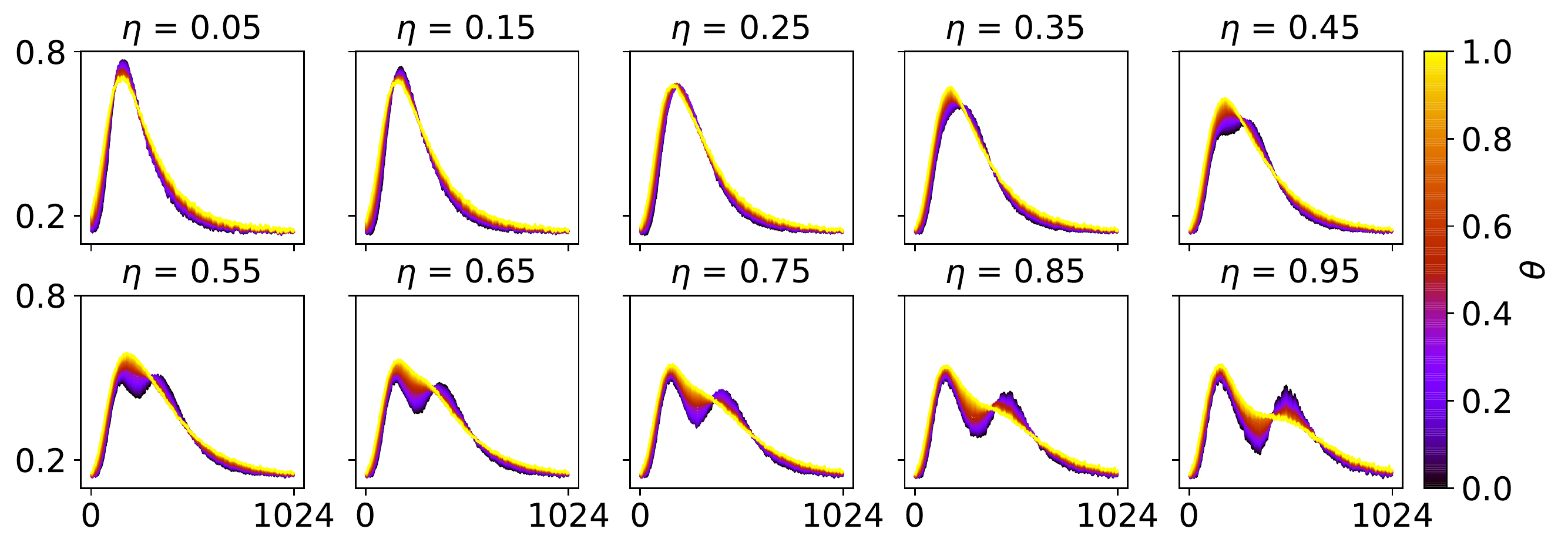}
\caption{\label{fig:sig_gen_rec}
Functional shape of $\hat{f}(\vec{p})$ as learned by the algorithm. A colormap of $\theta$ for a series of slices of $\eta$ is used to illustrate the dependency on both parameters.}
\end{figure}

\begin{figure}[ht!]
\centering
\includegraphics[width=.99\linewidth]{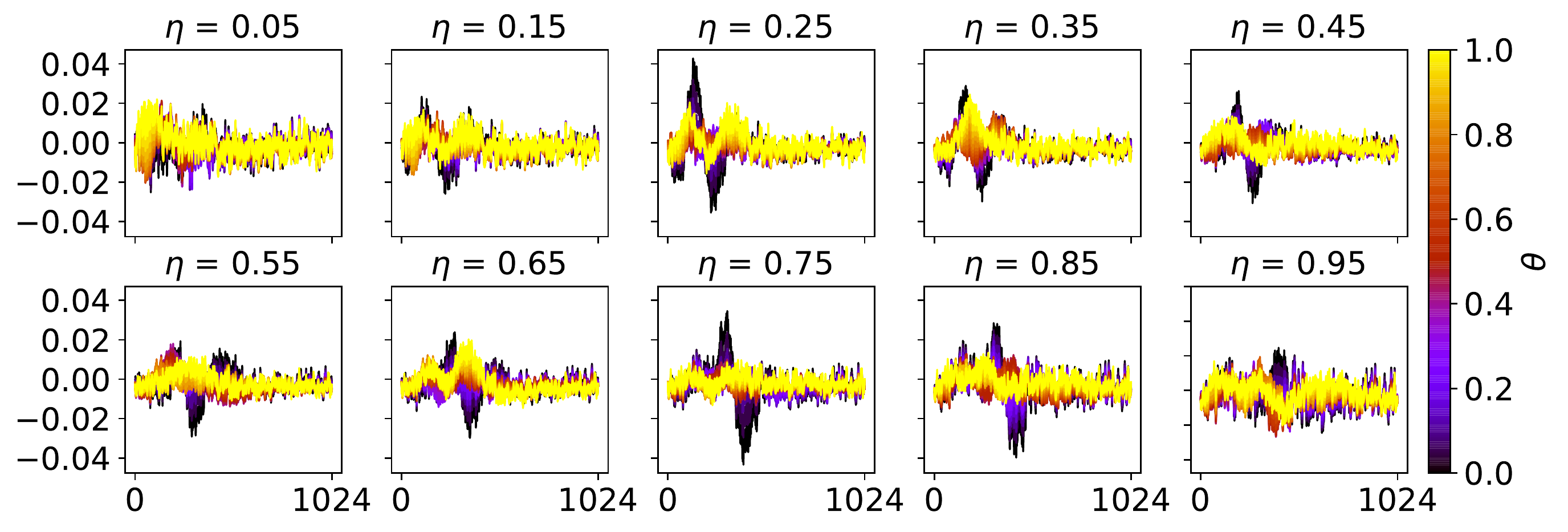}
\caption{\label{fig:sig_gen_dif} Subtraction of Figure~\ref{fig:signal_shape} and Figure~\ref{fig:sig_gen_rec} to illustrate the difference of $f(\vec{p})-\hat{f}(\vec{p})$.}
\end{figure}
\noindent The trained model can be used to characterize the transformation across the whole phase space, as presented in Figure~\ref{fig:sig_gen_rec}. In the ideal case, this scan should be equivalent to that earlier presented in Figure~\ref{fig:signal_shape}. Eye inspection allows to draw qualitative conclusions. In the first place, overall, the algorithm has learned a transformation $\hat{f}(\vec{p})$ directly from data examples which generally resembles the true one, and the predicted signals do not show worrisome artifacts or instabilities. This is remarkable given that the transformation is learned from noisy data examples but the algorithm directly provides a denoised output. To draw quantitative conclusions, in Figure~\ref{fig:sig_gen_dif} we present a plot of the difference of the true and the learned model, namely a plot of $f(\vec{p})-\hat{f}(\vec{p})$. For the majority of the phase-space the variations are very small, typically well contained within 2\% of the maximum amplitude. The largest differences are below 4\% of the maximum amplitude and are therefore sensibly smaller than the simulated noise level.

\subsection{Signal denoising}
\begin{figure}[ht!]
\centering 
\includegraphics[width=.99\linewidth]{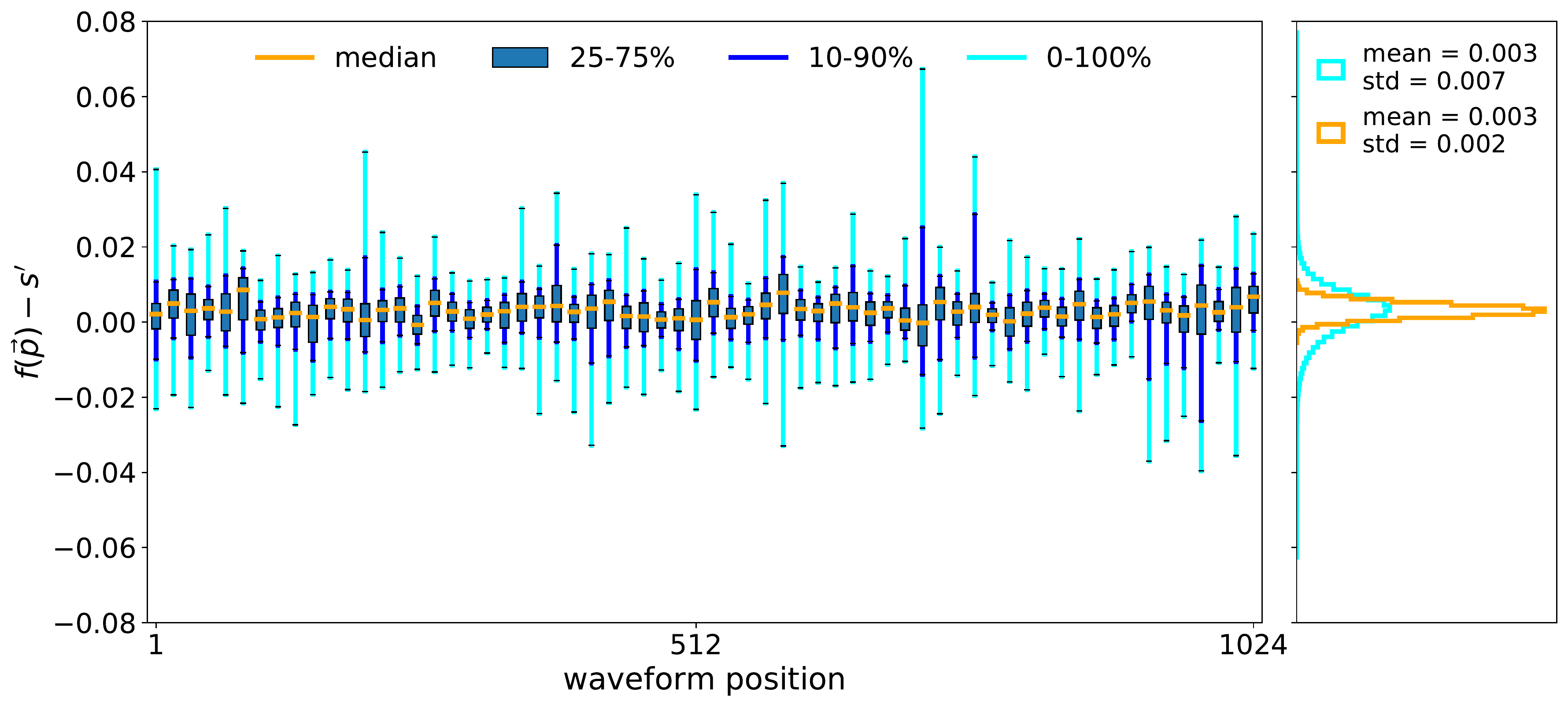}
\caption{\label{fig:signal_rec_err} Left panel: Distributions of the signal reconstruction error including various quantiles for one every sixteen waveform positions for the whole test dataset of 10000 waveforms. Right panel: Aggregated distribution of all 1024$\times$10000 signal reconstruction errors (cyan), and its 1024 medians (orange). The signal reconstruction error is defined as the difference between the normalized noiseless signals provided by Eq.~\ref{eq:f} and the signals generated by the transformation $\hat{f}^{-1}(\hat{f}(\vec{s}))\mapsto \vec{s'}$.}
\end{figure}
\begin{figure}[hpt!]
\centering 
\includegraphics[width=.90\linewidth]{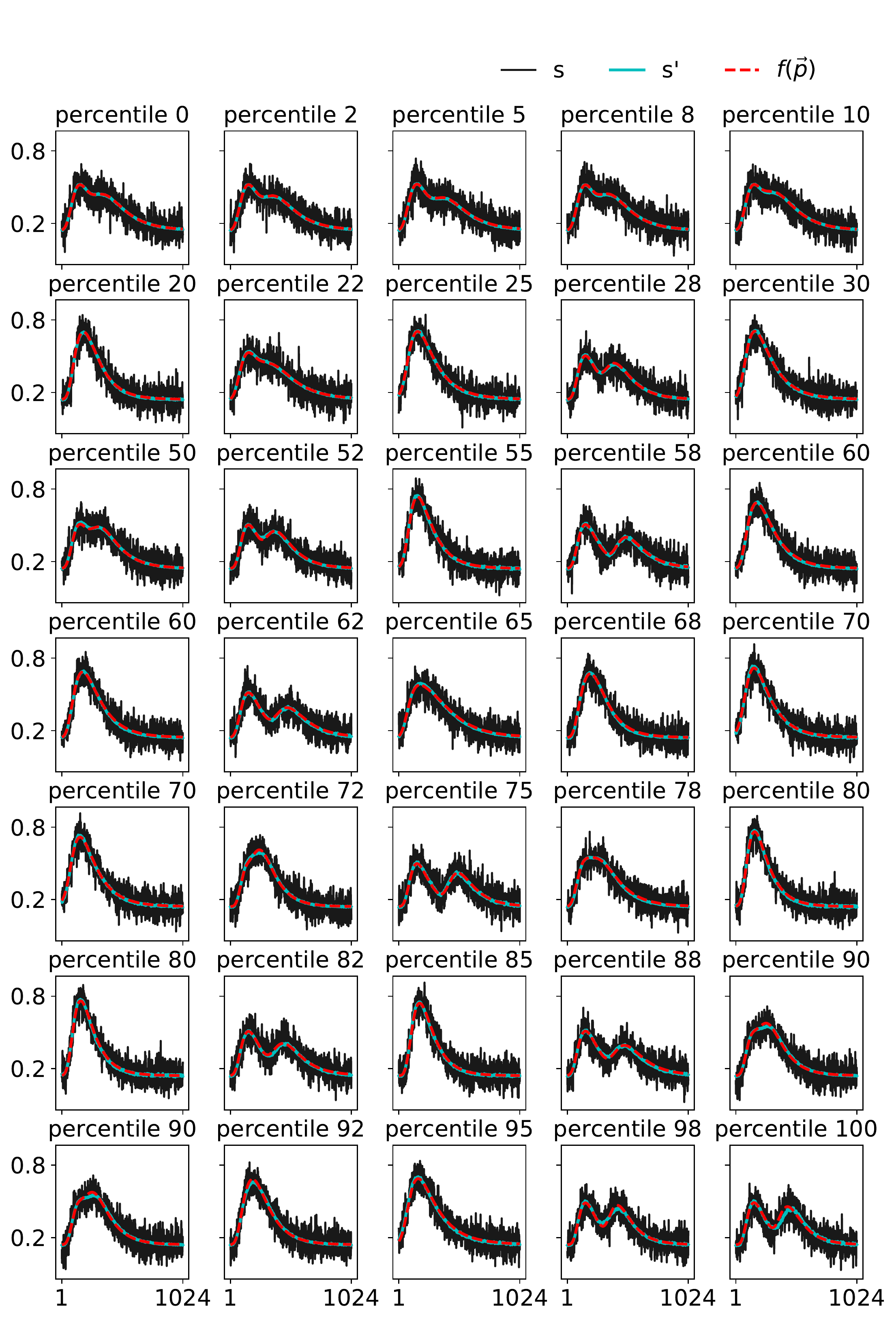}
\caption{\label{fig:reco_by_pct} The input noisy signal $\vec{s}$ is compared to the denoised generated signal $\vec{s'}$ and to the true shape of the toy model $f(\vec{p})$ for a total of 35 examples in the test dataset. The examples are chosen by their mean squared reconstruction error percentile. Percentile 0 (100) corresponds to the best (worse) reconstructed signal in the whole test dataset of 10k waveforms.}
\end{figure}
\noindent Finally, as we have learned both $\hat{f}(\vec{s})\mapsto\vec{p'}$ and $\hat{f}^{-1}(\vec{p})\mapsto \vec{s'}$ we can study the reconstruction of $\hat{f}^{-1}(\hat{f}(\vec{s}))\mapsto \vec{s'}$. This has two main advantages. On one hand the results of the composite transformation map noisy signals to denoised signals. In the second hand, the errors in the reconstruction of physical parameters and in the generation of signals are correlated, and in consequence the composite transformation might have a superior performance than the two individual transformations. \\
To evaluate the quality of the reconstructed denoised signals $\vec{s'}$ from noisy signals $\vec{s}$, the output of $\hat{f}^{-1}(\hat{f}(\vec{s}))\mapsto \vec{s'}$ has been compared to $f(\vec{p})$ for the 10k events in the test dataset. The results are presented in Figure~\ref{fig:signal_rec_err}. In the left of the Figure, the signal reconstruction error distributions for all waveform positions multiple of 16 are presented. The decision of showing only one in every 16 positions is driven by a clearer visualization, reducing the number of distributions from 1024 to 64. To show information for all the waveform positions, in the right part of the Figure aggregated results for all the 1024 positions are shown. These aggregated results consists of a distribution of the 1024 median positions in each waveform position (orange), and the 1024$\times$10$^4$ values reconstructed in any position for all events (cyan).
\\The results allow to draw numerous conclusions. In the first place, across all waveform positions the errors in the median are remarkably small, with a typical bias (standard deviation) similar to 0.3\% (0.2\%) of the maximum amplitude. The error in all reconstructed signals, is slightly wider but still of excellent quality with a typical bias of (standard deviation) 0.3\% (0.7\%). In the second place the reconstructed signals in all positions are densely packed around around the median without any significant outlier. The maximum error for the 64 positions under consideration is comparable with noise fluctuations and a very smooth and well controlled signal prediction by the network across the whole test dataset is observed.

\noindent To illustrate the accuracy and signal fidelity of the results presented in Figure~\ref{fig:signal_rec_err} all the 10k events in the test dataset have been sorted by their mean squared signal reconstruction error. Waveforms at different error percentiles are presented in Figure~\ref{fig:reco_by_pct}, being the percentile 0 (percentile 100) the event with minimum (maximum) signal reconstruction error. The results show excellent prediction fidelity across all percentiles with very minor differences even for percentile 100.

\section{Conclusions}
An algorithm consisting of a constrained bottleneck autoencoder has been presented as a reliable solution for the task of characterizing multidimensional signals directly from data. In particular, a toy problem has been presented and used to generate data. This data has been distorted by random noise and used to train and test the algorithm, blind to the toy model underlying functional form. Overall, the results show excellent performance in learning the true distribution mapping physical parameters into signals and the inverse of this distribution.\\
We have shown how to use the trained algorithm to do parameter estimations directly from signals and how to associate errors to this estimates. We have presented how, given some input physical parameters, the algorithm can be used to model signals with good precision and hence it is a suitable candidate to characterize the detector response in Monte Carlo simulations. Finally, we have shown the excellent potential of this algorithm in denoising signals.\\

Each of the former results open potential applications in HEP were the treatment and analysis of multidimensional signals, such as waveforms, is widely spread. In addition, the excellent accuracy this method in mapping signals to signals allows to envision compression applications in which the much lighter bottleneck representation is stored instead of the full signal.

\section{Acknowledgments}
C. Jes\'us-Valls acknowledges fruitful discussions with S.Alonso-Monsalve, C.Grimal-Bosch, S.~Bordoni and L.~Uboldi. C. Jes\'us-Valls and T. Lux acknowledge funding from the Spanish Ministerio de Econom\'{i}a y Competitividad (SEIDI-MINECO) under Grants No.~PID2019-107564GB-I00 and SEV-2016-0588. IFAE is partially funded by the CERCA program of the Generalitat de Catalunya. F. S\'anchez acknowledge the Swiss National Foundation Grant No. 200021\_85012.

\bibliographystyle{ieeetr}
\bibliography{bibliog}

\begin{thebibliography}{10}

\bibitem{moyal}
J.~Moyal, ``Xxx. theory of ionization fluctuations,'' {\em The London,
  Edinburgh, and Dublin Philosophical Magazine and Journal of Science},
  vol.~46, no.~374, pp.~263--280, 1955.

\bibitem{Korzenev:2021mny}
A.~Korzenev {\em et~al.}, ``{A $4\ensuremath{\pi}$ time-of-flight detector for
  the ND280/T2K upgrade},'' {\em JINST}, vol.~17, no.~01, p.~P01016, 2022.

\bibitem{ruthotto2021introduction}
L.~Ruthotto and E.~Haber, ``An introduction to deep generative modeling,'' {\em
  GAMM-Mitteilungen}, p.~e202100008, 2021.

\bibitem{DiGuglielmo:2021ide}
G.~Di~Guglielmo {\em et~al.}, ``{A reconfigurable neural network ASIC for
  detector front-end data compression at the HL-LHC},'' {\em IEEE Trans. Nucl.
  Sci.}, vol.~68, p.~2179, 2021.

\bibitem{Huang:2021ymz}
Y.~Huang, Y.~Ren, S.~Yoo, and J.~Huang, ``Efficient data compression for 3d
  sparse tpc via bicephalous convolutional autoencoder,'' in {\em 2021 20th
  IEEE International Conference on Machine Learning and Applications (ICMLA)},
  pp.~1094--1099, IEEE, 2021.

\bibitem{Yoon:2021mim}
B.~Yoon, C.~C. Chang, G.~T. Kenyon, N.~T.~T. Nguyen, and E.~Rrapaj,
  ``{Prediction and compression of lattice QCD data using machine learning
  algorithms on quantum annealer},'' {\em PoS}, vol.~LATTICE2021, p.~143, 2021.

\bibitem{Ai:2019yih}
P.~Ai, D.~Wang, G.~Huang, N.~Fang, D.~Xu, and F.~Zhang, ``{Timing and
  characterization of shaped pulses with MHz ADCs in a detector system: a
  comparative study and deep learning approach},'' {\em JINST}, vol.~14,
  no.~03, p.~P03002, 2019.
\newblock [Erratum: JINST 14, E03001 (2019)].

\bibitem{Shen:2019ohi}
H.~Shen, D.~George, E.~A. Huerta, and Z.~Zhao, ``Denoising gravitational waves
  with enhanced deep recurrent denoising auto-encoders,'' in {\em ICASSP 2019 -
  2019 IEEE International Conference on Acoustics, Speech and Signal Processing
  (ICASSP)}, pp.~3237--3241, 2019.

\bibitem{Erdmann:2019nie}
M.~Erdmann, F.~Schl\"uter, and R.~Smida, ``{Classification and Recovery of
  Radio Signals from Cosmic Ray Induced Air Showers with Deep Learning},'' {\em
  JINST}, vol.~14, no.~04, p.~P04005, 2019.

\bibitem{Farina:2018fyg}
M.~Farina, Y.~Nakai, and D.~Shih, ``{Searching for New Physics with Deep
  Autoencoders},'' {\em Phys. Rev. D}, vol.~101, no.~7, p.~075021, 2020.

\bibitem{Hajer:2018kqm}
J.~Hajer, Y.-Y. Li, T.~Liu, and H.~Wang, ``{Novelty Detection Meets Collider
  Physics},'' {\em Phys. Rev. D}, vol.~101, no.~7, p.~076015, 2020.

\bibitem{Blance:2019ibf}
A.~Blance, M.~Spannowsky, and P.~Waite, ``{Adversarially-trained autoencoders
  for robust unsupervised new physics searches},'' {\em JHEP}, vol.~10, p.~047,
  2019.

\bibitem{CrispimRomao:2020ucc}
M.~Crispim Rom\~ao, N.~F. Castro, and R.~Pedro, ``{Finding New Physics without
  learning about it: Anomaly Detection as a tool for Searches at Colliders},''
  {\em Eur. Phys. J. C}, vol.~81, no.~1, p.~27, 2021.
\newblock [Erratum: Eur.Phys.J.C 81, 1020 (2021)].

\bibitem{Cheng:2020dal}
T.~Cheng, J.-F. Arguin, J.~Leissner-Martin, J.~Pilette, and T.~Golling,
  ``{Variational Autoencoders for Anomalous Jet Tagging},'' {\em
  arXiv:2007.01850}, 7 2020.

\bibitem{Holl:2019xtt}
P.~Holl, L.~Hauertmann, B.~Majorovits, O.~Schulz, M.~Schuster, and A.~J.
  Zsigmond, ``{Deep learning based pulse shape discrimination for germanium
  detectors},'' {\em Eur. Phys. J. C}, vol.~79, no.~6, p.~450, 2019.

\bibitem{Liao:2021vec}
C.-H. Liao and F.-L. Lin, ``{Deep generative models of gravitational waveforms
  via conditional autoencoder},'' {\em Phys. Rev. D}, vol.~103, no.~12,
  p.~124051, 2021.

\bibitem{Monk:2018zsb}
J.~W. Monk, ``{Deep Learning as a Parton Shower},'' {\em JHEP}, vol.~12,
  p.~021, 2018.

\bibitem{voloshynovskiy2019information}
S.~Voloshynovskiy, M.~Kondah, S.~Rezaeifar, O.~Taran, T.~Holotyak, and D.~J.
  Rezende, ``Information bottleneck through variational glasses,'' {\em arXiv
  preprint arXiv:1912.00830}, 2019.

\bibitem{Buhmann:2020pmy}
E.~Buhmann, S.~Diefenbacher, E.~Eren, F.~Gaede, G.~Kasieczka, A.~Korol, and
  K.~Kr\"uger, ``{Getting High: High Fidelity Simulation of High Granularity
  Calorimeters with High Speed},'' {\em Comput. Softw. Big Sci.}, vol.~5,
  no.~1, p.~13, 2021.

\bibitem{Buhmann:2021vlp}
E.~Buhmann {\em et~al.}, ``{Fast and Accurate Electromagnetic and Hadronic
  Showers from Generative Models},'' {\em EPJ Web Conf.}, vol.~251, p.~03049,
  2021.

\bibitem{kingma2014adam}
D.~P. Kingma and J.~Ba, ``Adam: {A} method for stochastic optimization,'' in
  {\em 3rd International Conference on Learning Representations, {ICLR} 2015,
  San Diego, CA, USA, May 7-9, 2015, Conference Track Proceedings} (Y.~Bengio
  and Y.~LeCun, eds.), 2015.

\end{thebibliography}

\end{document}